\documentclass[10pt,A4paper,conference]{IEEEtran}
\usepackage{epsfig}
\usepackage{graphics}
\usepackage{amssymb}
\usepackage{cite}

\newcommand{\comment}[1]  {  }
\def\BE{\begin{equation}}
\def\EE{\end{equation}}
\def\BEA{\begin{eqnarray}}
\def\EEA{\end{eqnarray}}
\newcommand{\abs}[1]{\left\vert#1\right\vert}

\newcommand\mcX{{\mathcal{X}}}
\newcommand\mcY{{\mathcal{Y}}}

\newcommand\vd{{\bf d}}

\newcommand\vh{{\bf h}}

\newcommand\vq{{\bf q}}

\newcommand\vv{{\bf v}}

\newcommand\vx{{\bf x}}
\newcommand\vy{{\bf y}}

\newcommand\mI{{\bf I}}

\newcommand\mR{{\bf R}}
\newcommand\mS{{\bf S}}

\begin{document}

\title{On the Achievable Information Rates of
Finite-State Input Two-Dimensional Channels with Memory
}

\author{\authorblockN{Ori Shental}
\authorblockA{Dept. of Electrical Engineering-Systems\\
Tel Aviv University\\
Tel Aviv 69978, Israel \\
Email: shentalo@eng.tau.ac.il}
\and
\authorblockN{Noam Shental}
\authorblockA{Dept. of Physics of Complex Systems\\
Weizmann Institute of Science\\
Rehovot 76100, Israel\\
Email: fenoam@wisemail.weizmann.ac.il} \and
\authorblockN{Shlomo Shamai (Shitz)}
\authorblockA{Dept. of Electrical Engineering\\
Technion-Israel Institute of Technology\\
Haifa 32000, Israel\\
Email: sshlomo@ee.technion.ac.il}
}

\maketitle

\begin{abstract}

The achievable information rate of finite-state input two-dimensional (2-D)
channels with memory is an open problem, which is
relevant, e.g., for inter-symbol-interference (ISI) channels and
cellular multiple-access channels.
We propose a method for simulation-based computation
of such information rates.
We first draw a connection between the Shannon-theoretic information rate and
the statistical mechanics notion of free energy.
Since the free energy of such systems is intractable,
we approximate it using the cluster variation method,
implemented  via generalized belief propagation.
The derived, fully tractable, algorithm is shown to
provide a practically accurate estimate of the information rate.
In our experimental study we calculate the information rates of
2-D ISI channels and of hexagonal Wyner cellular networks with binary inputs,
for which formerly only bounds were  known.
\end{abstract}
\rule{0mm}{0mm}
                \mbox{}
                \mbox{}\hrulefill%
                   {{\sf Submitted to ISIT 2005}}

\section{introduction}
Two-dimensional (2-D) finite-state input channels with memory
exhibit an important class of channels, which appears extensively
in a wide range of fields.
For example, in inter-symbol interference (ISI) channels,
which are applicable to magnetic and optical recording devices,  finite-state
symbols are ordered on a 2-D grid, causing interference in a limited neighborhood.

A second example concerns multiple-access  channels in cellular networks.
In a seminal work~\cite{BibDB:Wyner}, Wyner has introduced a simple,
yet insightful, analytically solvable model for cellular
networks with Gaussian signaling,
thus yielding a considerable insight into the ultimate information-theoretic limits of
realistic cellular networks. In addition to a naive one-dimensional (1-D) extension of  a single cell system,
Wyner has also analyzed the traditional 2-D hexagonal topology, where interference is caused by neighboring
cellular tiers. Hence, in case of binary signaling Wyner's model can be viewed as an instance of a finite-state
input dispersive channel.

The  capacity of finite-state input dispersive  channels is defined
as the maximum mutual information rate over all input distributions.
Computing this capacity for 1-D and 2-D channels is  an open problem.
Calculating the mutual information rate in the  case  of
a predefined stationary input distribution is, in principle, a simpler problem.
For example, for  input symbols which are i.i.d. and equiprobable,
this is termed  the symmetric information rate (SIR),
thus providing a limit on the achievable rate of reliable communication
in this common~case.

Various bounds, either
rigorous~\cite{BibDB:PhDHirt,BibDB:ShamaiOzarowWyner,BibDB:ChenSiegelMarkovIT},
numerical~\cite{BibDB:Kavcic,BibDB:YangKavcic} or
conjectured~\cite{BibDB:ShamaiLaroia}, on the capacity and SIR of
certain finite-state input 1-D dispersive channels have been
proposed. Recently several authors introduced simulation based
methodologies for  computing  such information
rates (\cite{BibDB:ArnoldLoeligerIT} and references therein). In
this approach, the forward recursion of the sum-product (BCJR)
algorithm~\cite{BibDB:BCJR} is used for estimating the
a-posteriori probability (APP) and consequently deriving the 1-D
information rates. As for 2-D channels, due to their  inherent
complexity, only upper and lower bounds on the information rate
are known~\cite{BibDB:ChenSiegel}.

In this paper we propose a simulation-based method for estimating the
information rate of 2-D channels. This method can be viewed as an extension
of its 1-D Monte-Carlo counterpart~\cite{BibDB:ArnoldLoeligerIT}, where a fully tractable
generalized belief propagation (GBP) receiver replaces the sum-product algorithm as an APP
 inference engine.

In a former work~\cite{BibDB:Shental} we have shown that  a GBP receiver serves excellently
well as an APP detector of dispersive 2-D channels\footnote{A detector which is based
on standard belief propagation often fails to converge in 2-D channels.}.
In this work we utilize another aspect of GBP, i.e., its remarkable
ability to  approximate the free energy of 2-D channels, as we draw the
connection between the information rate and the free energy~\cite{BibDB:Tanaka}.

The  paper is organized as follows. Section~\ref{sec:channel model} introduces the
dispersive 2-D channel model, while section~\ref{sec:inf rate}
derives the form of the information rate and draws its connection to the free energy.
Since the free energy of 2-D channels is intractable,   section~\ref{sec:CVM}
presents a  method for approximating it, which is then applied
in the context of probabilistic graphical models in section~\ref{sec:graphical models}.
Section~\ref{sec:results} evaluates  the quality of the free energy approximation, as compared
to its exact value. Next, simulation results for the information rate of a 2-D ISI channel
and an hexagonal Wyner cellular network are provided. The results  are discussed in
section~\ref{sec:discussion}.

We shall use the following notations. The operator $\{\cdot\}^{T}$ stands for a vector or matrix transpose,
$\{\cdot\}_{i}$ and $\{\cdot\}_{ij}$ denote entries of a vector and matrix, respectively.

\section{channel model}
\label{sec:channel model}
Consider a $N\times N$ 2-D finite-state input channel with memory in
the form
\BE
\label{eq_channel}
y_{k,l}=d_{k,l}+v_{k,l}+\sum_{(i,j)\in \langle k,l\rangle}\alpha_{i,j}d_{i,j}\quad\forall k,l=1,\ldots,N,
\EE
where $y_{k,l}$, the channel's output observation at symbol $(k,l)\in\mathbb{Z}^{2}$,
is the sum of the finite-state alphabet input symbol $d_{k,l}$,
assumed to be taken from a stationary process, and two additional terms.
The first term $v_{k,l}$ represents ambient additive white Gaussian noise (AWGN),
while the second term is the scaled interference caused by adjacent
symbols to $(k,l)$, denoted by $\langle k,l\rangle$.
The parameter $\alpha_{i,j}$ ($\abs{\alpha_{i,j}}\leq 1$) controls the interference attenuation.
The interference term is assumed to be spatially invariant (excluding boundary symbols),
which together with assumptions regarding $d_{i,j}$ and $v_{i,j}$  guaranties that
$y_{k,l}$ ($k,l=1,\ldots,N$) are stationary random variables.
We also assume that the channel is perfectly known on the receiver's side,
which can jointly process all observations.

\begin{figure}[thb!]
\begin{center}
\begin{tabular}{cc}
\psfig{file=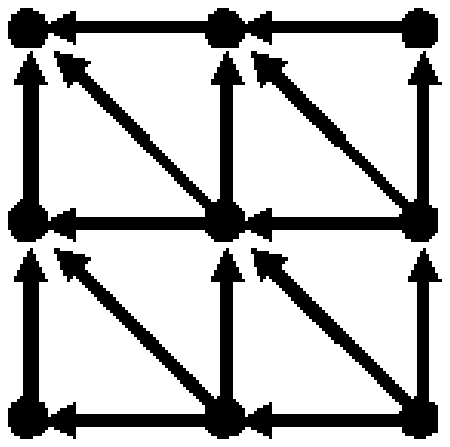,width=0.13\textwidth}&
\psfig{file=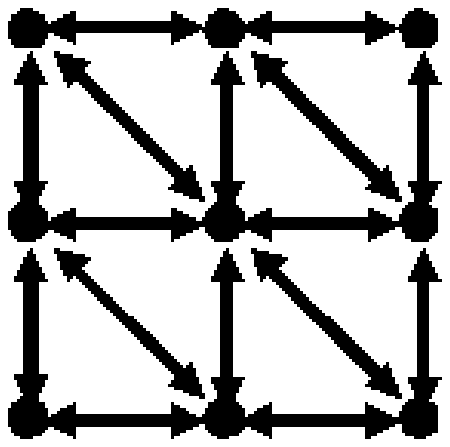,width=0.13\textwidth}
\\(a) & (b)\\
\psfig{file=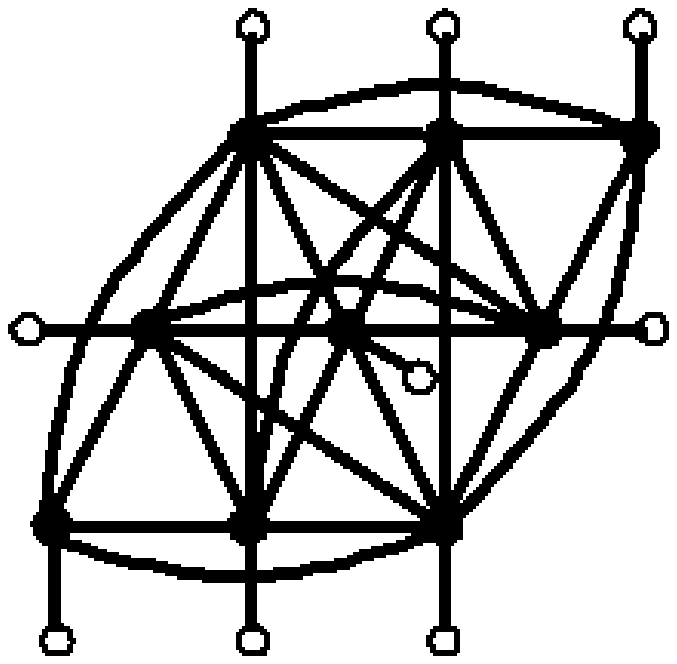,width=0.14\textwidth} &
\psfig{file=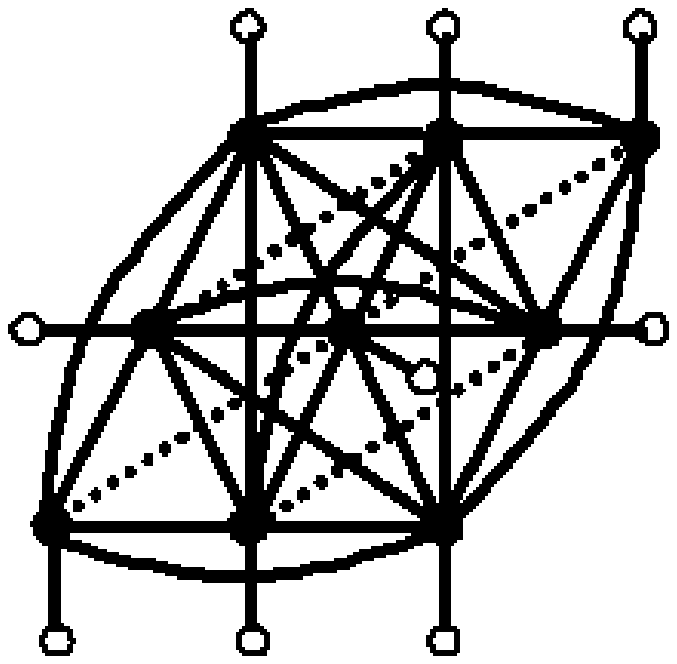,width=0.14\textwidth}
\\(c) & (d)
\end{tabular}
\end{center}
\caption{{\bf Upper pane:} Interference structures for two types of $3\times 3$ 2-D channels:
(a) ISI grid, (b) hexagonal Wyner cellular network. The arrows mark the direction of
interference. {\bf Lower pane:} The corresponding undirected graphical model representation of the channels in
the upper pane: (c) ISI grid, (d) hexagonal Wyner cellular network. Full nodes represent
(hidden) transmitted bits, while empty nodes correspond to the observations. Interaction couplings
(compatibility function) $\psi_{ij}$ are denoted by a solid line connecting two full nodes, while the external
field potential (evidence) $\phi_{i}$ is depicted by a solid line connecting a full node and an empty node. For clarity we
use dotted edges in (d) to represent the extra edges added compared to the graph (c).}
\label{fig_topology_and_MRF}
\end{figure}

Stacking all the observations, data symbols and noise samples into $N^{2}\times 1$ vectors $\vy$,
$\vd$ and $\vv$, respectively, (\ref{eq_channel}) can be rewritten as
\BE
\label{eq_vectorModel}
\vy=\mS\vd+\vv,
\EE
where the $N^{2}\times N^{2}$ matrix $\mS$ encapsulates the memory/interference structure.
Each 2-D channel is uniquely defined by its interference matrix $\mS$.
Our basic assumption, which later allows for  a graphical model interpretation,
is that interference is caused by neighboring symbols, i.e., $\mS$ is  a relatively sparse matrix.
The upper pane in
Fig.~\ref{fig_topology_and_MRF} represents the interference structure of two topologies: ISI (a)
and an hexagonal Wyner cellular network (b).
In the following derivations we assume real-space data signaling $\vd$, interference $\mS$ and noise
$\vv\sim\mathcal{N}(\mathbf{0},\sigma^{2}\mI_{N})$
(an extension to the complex domain is straightforward.)

\section{Information rate}
\label{sec:inf rate}
\subsection{Basic Definitions}
\label{sec:inf rate basic def}
The information rate,  i.e. mutual information per symbol,
between the channel's input $\mathcal{X}$ and output $\mathcal{Y}$ is,
\begin{eqnarray}
\label{eq_I}
I(\mathcal{X};\mathcal{Y})=h(\mathcal{Y})-h(\mathcal{Y|X}),
\end{eqnarray}
where $h(\cdot)$ are (differential) entropy rates, where, by definition, the entropy  rate $h(\mathcal{Q})$
of a stationary process $\vq=\{q_{1},\ldots,q_{L}\}^{T}$ is given by $\lim_{L\rightarrow \infty} h(\vq)/L$.
Let us deal separately with the two terms in~(\ref{eq_I}).

The second term, $h(\mathcal{Y|X})$, is given by
$\lim_{N\rightarrow \infty} h(\vy|\vx)/N^{2}$, but since $h(\vy|\vx)=h(\vv)$,
and $\vv$ is AWGN, it is straightforward to validate that $h(\mathcal{Y|X})=(\log{2\pi e\sigma^{2}})/2$.

In order to calculate $h(\mathcal{Y})$ we apply the Shannon-McMillan-Breiman
theorem~\cite{BibDB:BookCover}\footnote{The theorem also applies to continuous random variables~\cite{BibDB:Leroux}.},
which states  that for a stationary and ergodic channel the entropy rate can be calculated by
\BE
\label{eq_SMB}
-\frac{1}{N^{2}}\log p(\vy)\stackrel{N \rightarrow \infty}{\longrightarrow}
 h(\mcY) ~~\mbox{with probability}~~ 1,
\EE
where $p(\vy)$ is the joint distribution of the channel's output $\vy$.
Hence, in order to calculate  the information rate one needs to calculate $p(\vy)$
in the limit of large systems, as described in the next section.

\subsection{The Connection  to Free Energy}
\label{inf rate free energy}
Using Bayes' law $p(\vy)$ can be rewritten as
\BE
\label{eq_py}
    p(\vy)=\sum_{\vx}p(\vy|\vx)\Pr(\vx)=\sum_{\vx}p(\vv)\Pr(\vx),
\EE
where $\sum_{\vx}$ corresponds to a sum over all the possible values of the transmitted symbols $\vd$.
Hereinafter, for exposition purposes we consider the case of equiprobable i.i.d.
binary-input alphabet, i.e., $d_{i}\in \pm 1$.
Hence, using the distribution of $\vv$, (\ref{eq_py}) can be rewritten as
\BE
\label{eq_C/Z}
    p(\vy)=\mathcal{Z}\cdot(2\mathcal{C})^{-N^{2}},
\EE
where $\mathcal{C}\triangleq(2\pi\sigma^{2})^{1/2}$, and
\BE
    \mathcal{Z}\triangleq\sum_{\vx}\exp{\bigg(-\frac{1}{2\sigma^2}||\vy-\mS\vx||^{2}\bigg)},
\EE
is the {\em partition function}.

Inserting (\ref{eq_C/Z}) into (\ref{eq_SMB}), the $I(\mathcal{X};\mathcal{Y})$  can be written as
\BE
\label{eq_inf_rate}
\log{2}-1/2+\mathcal{F}
\stackrel{N \rightarrow \infty}{\longrightarrow}
I(\mcX;\mcY)
~~\mbox{with probability}~~ 1,
\EE
where
\BE
\label{eq_free_energy}
\mathcal{F}\triangleq -\frac{1}{N^{2}}\log(\mathcal{Z})
\EE
is recognized as the free energy per symbol~\cite{BibDB:BookMezardEtAl,BibDB:Tanaka}.
Hence, the problem of calculating the information rate boils down to
estimating the free energy of an infinite system, as discussed in the next section.
The information rate in (\ref{eq_inf_rate}) is termed the {\em symmetric}
(a.k.a. uniform-input) information rate (SIR),
due to the assumption regarding the uniformity of the input symbols.
Similar analysis also holds for other stationary finite-state input distributions.

\section{The Free Energy and the Cluster Variation Method}
\label{sec:CVM}
The free energy is a fundamental quantity in statistical mechanics which
the physics literature has devoted a considerable effort in calculating.
However, evaluating the free energy of infinitely
large 2-D channels such as~(\ref{eq_channel}) is infeasible
and, one  must resort to approximate methods\footnote{Our system corresponds to  a random field 2-D
Ising system, for which an analytical  solution is not available~\cite{BibDB:BookMezardEtAl}.}.

One of the classic approximation methods of free energies is the
Kikuchi approximation, also known as the {\em cluster variation method} (CVM,~\cite{BibDB:ReportYedidiaEtAl}).
The difficulty in exactly calculating  the free energy  results from the intractability of the
probability distribution $p(\vy)$.
Hence, the CVM follows a variational principle:
It defines the free energy as a functional of this probability distribution,
$\mathcal{F}(p(\vy))$, replaces $p(\vy)$ by a tractable trial belief vector
$b(\vy)$
\footnote{The trial belief vector $b(\vy)\triangleq \prod_{\lambda\in M} p(\vy_{\lambda})^{c_{\lambda}}$,
where $\lambda$ is a `cluster' of neighboring symbols $\vy_{\lambda}$, taken from the set `clusters' $M$.
The integers  $c_{\lambda}$, a.k.a. {\em counting numbers}, are provided by the CVM
in order to ascertain that each symbol is counted exactly once in the corresponding free energy.
Since  $b(\vy)$ depends only on local marginal probabilities, $p(\vy_{\lambda})$,  it is tractable.
However, $b(\vy)$   need not necessarily form a valid probability distribution function~\cite{BibDB:ReportYedidiaEtAl}.},
then minimizes $\mathcal{F}(b(\vy))$
w.r.t $b(\vy)$, and considers the minimal value
as its approximation to the free energy.
Hence, our idea for estimating the information rate is to use the CVM over a large enough,
yet finite, system, as  the computed free energy per symbol is conjectured to converge to its
exact value for infinite systems.
This idea is empirically  validated   in section~\ref{sec:results}.

Recently, Yedidia et al.~\cite{BibDB:ReportYedidiaEtAl} have proved a correspondence between
the stationary points  of the CVM-based free energy and the fixed points of
a message passing algorithm from the field of graphical models termed {\em generalized belief propagation} (GBP).
GBP is an extension of the celebrated belief propagation algorithm (BP),
that has been shown to provide better approximations than BP.
Note, in passing,  that  Yedidia et al., have also shown that the fixed points of BP
correspond to the stationary points of the Bethe free energy, which is a special
case of the CVM, in the same way that BP is a special case of GBP.
For an elaborate   discussion of both CVM and GBP see~\cite{BibDB:ReportYedidiaEtAl}.
In the following section we describe the channel  from
the perspective of graphical models, and shortly describe our application of the GBP algorithm.

\section{The connection to undirected graphical models}
\label{sec:graphical models}
An undirected graphical model with pairwise potentials  (a.k.a. pairwise Markov random fields),
consists of a graph $G$ and potential (compatibility) functions $\psi_{ij}(\tilde{x_{i}},\tilde{x_{j}})$
and $\phi_{i}(\tilde{x_{i}})$
such that the probability of an assignment $\tilde{\vx}$ is given by
\BE
\label{eq_graph}
    \Pr(\tilde{\vx}) \propto \prod_{(i>j)}{\psi_{ij}(\tilde{x_i},\tilde{x_j})}\prod_{i}{\phi_i(\tilde{x_i})}.
\EE
The notation $(i>j)$ represents the set of all connected pairs $(\tilde{x_i},\tilde{x_j})$.

The joint posterior probability of the channel can be written as
\BE
\label{eq_posterior}
\Pr(\vx|\vy)=\mathcal{Z}^{-1}\exp{\bigg(-\frac{1}{2\sigma^2}||\vy-\mS\vx||^{2}\bigg)}.
\EE
Hence (\ref{eq_posterior}) defines the  undirected graphical model
\BE
    \Pr(\vx|\vy) \propto\prod_{(i>j)}{\psi_{ij}(x_i,x_j)}\prod_{i}{\phi_i(x_i,h_i)},
\EE
where
\BE
    \psi_{ij}(x_i,x_j)=\exp\big({-\frac{R_{ij}x_{i}x_{j}}{\sigma^{2}}}\big)
\EE is a compatibility function representing the structure of the system and the potential
\BE
    \phi_i(x_i,y_i)=\exp\big({\frac{h_{i}x_{i}}{\sigma^{2}}}\big)
\EE is the 'evidence' or local likelihood, which describes the statistical dependency between the hidden
variable $x_i$ and the observed variable $h_i$\footnote{Notice that for the non-binary finite-state input
alphabet case, the Markov random fields  modelling  is identical, except for an additional external field potential operating on
each node which can be absorbed into $\phi_{i}$ term. This additional potential arises from the
auto-correlations $R_{ii}$, which can not be dropped out from the sufficient statistics expression as in the
binary case.}. The matrix $\mR=\mS^{T}\mS$ is the interference cross-correlation matrix and
$\vh=\mS^{T}\vy$ is the output vector of a filter matched to the channel's interference structure.
The lower pane in Fig.~\ref{fig_topology_and_MRF} presents the resulting graphical models of the two
channel examples considered in this work.

\subsection{Generalized Belief Propagation}
The GBP algorithm  is an extension of BP that
has been shown to provide better approximations in many applications.
The first step in applying GBP to a  graph (\ref{eq_graph}) is to define regions (clusters)
of nodes which may intersect, and then pass messages between these regions in an analogous way to BP.
Within each such region GBP performs {\em exact} inference, thus short cycles of nodes
which are included in a region cause no problem.

\begin{figure}
\parbox{3.5cm}{
\begin{tabular}{c}
\psfig{file=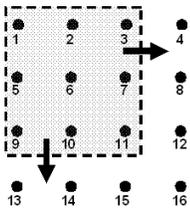,width=0.15\textwidth}
\end{tabular}
} \hfill
\parbox{5cm}{\caption{Covering a $4\times 4$ 2-D channel by $3\times 3$ regions used in GBP. Regions are
defined by sliding a $3\times 3$ window along the channel. The result is four regions for this $4\times 4$
system:\newline
 $\{1,2,3,5,6,7,9,10,11\}$,
 $\{2,3,4,6,7,8,10,11,12\}$,
 $\{5,6,7,9,10,11,13,14,15\}$,
$\{6,7,8,10,11,12,14,15,16\}$.}
\label{fig_slidingwindow_and_regiongraph}}
\end{figure}

Hence, a region that encompasses all nodes along the shortest cycles, might be a
desired choice. Since the graphical models of our 2-D channel examples contain
interactions between nearest neighbors and  next nearest neighbors,
as displayed in Fig.~\ref{fig_topology_and_MRF}-(c,d), a natural choice of regions
is a sliding $3\times 3$ square of nodes (e.g., see Fig~\ref{fig_slidingwindow_and_regiongraph}).
In all of our simulations the selected GBP regions were of size $3\times 3$.
Surprisingly, the computations required for GBP are only slightly larger than the computations required for BP,
and its complexity grows exponentially only with the size of the chosen regions.

\section{Simulation Results}
\label{sec:results}
\subsection{Quality of the Free Energy Approximation}
\begin{figure}[thb!]
\begin{center}
\psfig{file=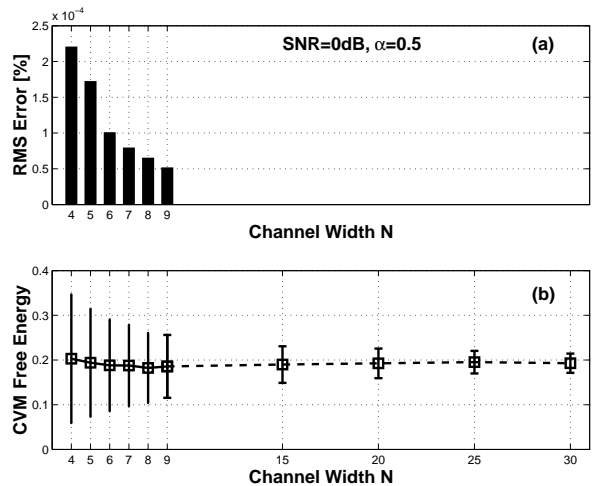,width=0.43\textwidth}
\end{center}
\caption{(a) Root mean square (RMS) error (in \%) in computing the free energy per symbol,
exactly and using the CVM, for $N\times N$ channels.
The results were obtained using  500 realizations of
 Wyner's hexagonal cellular networks (assuming a single user per cell),
 with  $\alpha=0.5$ and SNR=$0$dB.
(b) The corresponding  CVM  free energy per symbol as a function of $N$.
For $N\leq 9$ we used the same realizations as in (a). For larger systems (dashed line)
the exact free energy can not be calculated. }
\label{fig_conv}
\end{figure}

In order to evaluate the quality of our free energy approximation,
we performed  Monte-Carlo simulations of several channels.
Fig.~\ref{fig_conv}-(a) displays the root mean square (RMS) error per symbol,  in percentage,
between the approximated and  exact free energies
as a function of the  channel's size $N^{2}$
($N=4,\ldots,9$, where a $9\times 9$ channel is the largest
case for which exact computation was feasible.)
The results were averaged over 500 realizations.
As  can be observed the difference between the approximated and exact free energies is minuscule
(in the order of $10^{-4}\%$).
These results were obtained for a specific channel,
i.e., Wyner's hexagonal cellular network, as depicted in
Fig~\ref{fig_topology_and_MRF}-(b), with  $\alpha=0.5$ and signal to noise ratio (SNR) of $0$dB.
 Similar error performance was  observed for all other channels, throughout the entire interference range
and for a wide scope of SNR.

Fig.~\ref{fig_conv}-(b) presents the CVM approximation of the free energy per symbol
as a function of the channel's size.
The  results were averaged over 500 channel realizations
(for small channel size, $N\leq9$, we used  the same channel
realizations as   in Fig~\ref{fig_conv}-(a).)
It can be observed that the free energy per symbol converges with the size of the system,
and that the  differences among  realizations become smaller.
In principle,  we could have simulated even larger systems,
for which these differences would have been smaller.
However,  it seems that a $30\times 30$ system size suffices
as an approximation of the exact free energy per symbol of infinite systems,
thus can provide   a proper estimate  of the information rate.

\subsection{Information Rate Computation}
The proposed GBP-based algorithm is used for estimating the SIR of two examples of dispersive 2-D
channels: a 2-D ISI channel and an hexagonal Wyner cellular network.
The results were obtained by averaging over $1000$ realizations of   $30\times30$ channels.
The standard deviation of  the  results
were small, thus are omitted from the figures.

\begin{figure}[thb!]
\begin{center}
\psfig{file=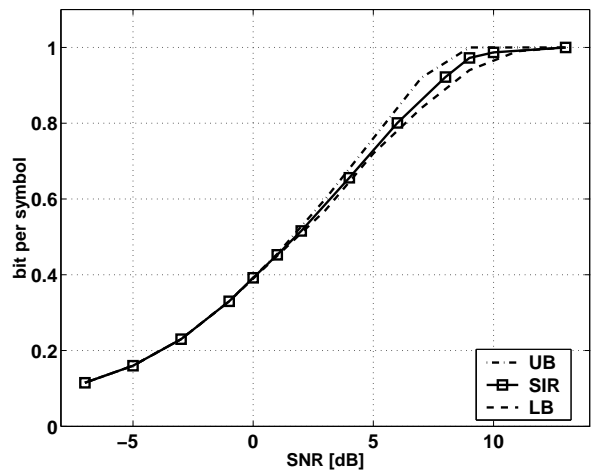,width=0.43\textwidth}
\end{center}
\caption{A 2-D ISI channel: SIR, in terms of bit per symbol,  evaluated using GBP-based simulations
(squares and a solid line), as a function of SNR.
Also shown are     upper (UB, dashed-dotted) and lower (LB, dashed) bounds on the SIR~\cite{BibDB:ChenSiegel}.
}\label{fig_ISI}
\end{figure}

\paragraph{2-D ISI Channel}
We compute the SIR of a binary ISI channel with non-trivial ($\alpha=0.5$) interference structure as depicted in
Fig.~\ref{fig_topology_and_MRF}-(a).
Fig.~\ref{fig_ISI} presents the SIR, in terms of bit per symbol,  computed using the
GBP-based algorithm, as a function of SNR. Also drawn are the lower and upper bounds on the SIR,
recently suggested by Chen and Siegel~\cite{BibDB:ChenSiegel}.
As can be seen the evaluated SIR  agrees with these tight  bounds.

\begin{figure}[thb!]
\begin{center}
\psfig{file=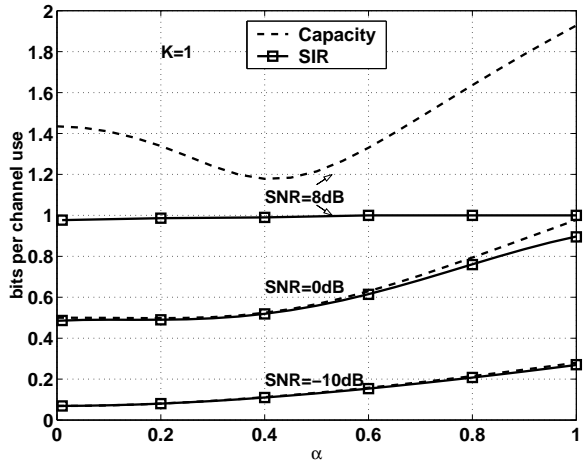,width=0.43\textwidth}
\end{center}
\caption{Hexagonal Wyner network: SIR (squares and solid line) and Gaussian signaling capacity (dashed),
in bits per channel use, as a function of  $\alpha$ for three  SNR values: $-10$, $0$ and $8$ dB, and
a single user within each cell ($K=1$).}
\label{fig_Wyner}
\end{figure}

\paragraph{2-D Wyner Cellular Network}
In a similar way, we computed the SIR of an hexagonal topology Wyner
model~\cite{BibDB:Wyner}, under binary signaling,  with a single user within each cell (i.e. $K=1$ in Wyner's notation).
Fig.~\ref{fig_Wyner} displays the SIR calculated for the possible range of inter-cell interference scaling
$\alpha$, for three SNR levels. For comparison we also present the Gaussian signaling capacity,
as derived by Wyner.
As may  be expected for low SNR ($-10$dB) the SIR and Wyner's capacity almost coincide.
For the intermediate  SNR level ($0$dB) Wyner's capacity provides a tight upper bound on the SIR
for $\alpha<0.5$. Note, in passing, that since the capacity of a binary channel is bounded
between the SIR and Wyner's Gaussian capacity, one can also infer  the capacity
in these low and intermediate SNR regimes.
As for  high SNR ($8$dB) the SIR saturates the $1$-bit bound, for almost all values of $\alpha$.

\begin{figure}[thb!]
\begin{center}
\psfig{file=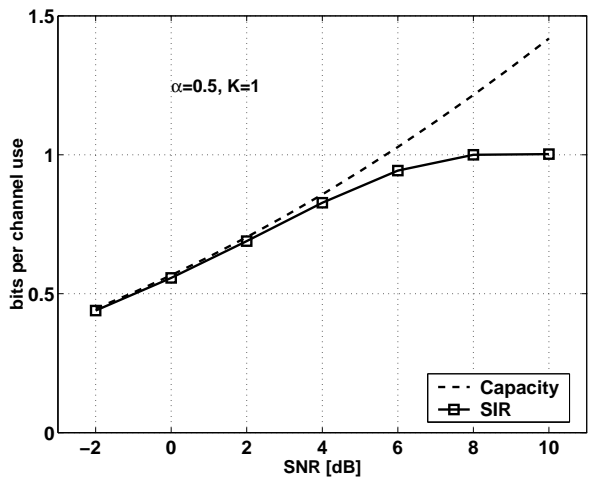,width=0.43\textwidth}
\end{center}
\caption{Hexagonal Wyner network: SIR (squares and solid line) and capacity (dashed), in bits per channel use,
as a function of SNR for $\alpha=0.5$ and $K=1$.
}
\label{fig_SNR}
\end{figure}

In Fig.~\ref{fig_SNR} we  evaluated the SIR for a fixed  $\alpha=0.5$ as a function
of SNR.
The SIR coincides with Wyner's capacity  for  $\mbox{SNR}\lesssim 4$dB.

It should be emphasized that the  analysis could have been performed,
in a similar manner, for the case of several intra-cell users, i.e. $K>1$.
This  case corresponds to the case of $K=1$, where ($K+1$)-ary signaling
from a binomial distribution, replaces the equiprobable binary signaling.

\section{Discussion}
\label{sec:discussion}
In this paper we introduced a method for a simulation-based computation of the information rates of
2-D finite-state input channels with memory.
Our method is established upon a connection
between the information rate and the free energy,
and on a graphical models based  method of approximating this free energy.
The quality of the approximation was compared to the exact free energy  using small channels,
and was  found to exhibit practically accurate behavior,
being consistent both as a function of the SNR and over the possible interference range.
This behavior is then conjectured to hold for large, yet finite, systems,
for which the information rate was estimated.
In order to validate our  method we compared the resulting information rate to formerly
calculated bounds.

The physics and graphical models literature does not provide a rigorous explanation
for this  remarkable quality of approximation, as provided by the GBP-based CVM,
thus research in this direction is currently underway.

\section*{Acknowledgment}
The authors are grateful to Ido Kanter and to Yair Weiss for useful discussions,
and to Dongning Guo for constructive comments.

\bibliographystyle{IEEEtran}
\bibliography{isit}

\begin{thebibliography}{10}
\providecommand{\url}[1]{#1}
\def\UrlFont{\rmfamily}
\providecommand{\newblock}{\relax}
\providecommand{\bibinfo}[2]{#2}
\providecommand\BIBentrySTDinterwordspacing{\spaceskip=0pt\relax}
\providecommand\BIBentryALTinterwordstretchfactor{4}
\providecommand\BIBentryALTinterwordspacing{\spaceskip=\fontdimen2\font plus
\BIBentryALTinterwordstretchfactor\fontdimen3\font minus
  \fontdimen4\font\relax}
\providecommand\BIBforeignlanguage[2]{{%
\expandafter\ifx\csname l@#1\endcsname\relax
\typeout{** WARNING: IEEEtran.bst: No hyphenation pattern has been}%
\typeout{** loaded for the language `#1'. Using the pattern for}%
\typeout{** the default language instead.}%
\else
\language=\csname l@#1\endcsname
\fi
#2}}

\bibitem{BibDB:Wyner}
A.~D. Wyner, ``Shannon-theoretic approach to a gaussian cellular
  multiple-access channel,'' vol.~40, pp. 1713--1727, Nov. 1994.

\bibitem{BibDB:PhDHirt}
W.~Hirt, ``Capacity and information rates of discrete-time channels with
  memory,'' Ph.D. dissertation, Swiss Federal Inst. of Tech. (ETH), Zurich,
  Switzerland, 1998.

\bibitem{BibDB:ShamaiOzarowWyner}
\mbox{S. Shamai (Shitz)}, L.~H. Ozarow, and A.~D. Wyner, ``Information rates
  for a discrete-time gaussian channel with intersymbol interference and
  stationary inputs,'' vol.~37, no.~6, pp. 1527--1539, Nov. 1991.

\bibitem{BibDB:ChenSiegelMarkovIT}
J.~Chen and P.~H. Siegel, ``Markov processes asymptotically achieve the
  capacity of finite state intersymbol interference channels,'' To Appear.

\bibitem{BibDB:Kavcic}
A.~Kav\v{c}i\'{c}, ``On the capacity of markov sources over noisy channels,''
  in \emph{Proc. {IEEE} Global Conference on Communications ({GLOBECOM})}, San
  Antonio, Texas, USA, Nov. 2001, pp. 2997--3001.

\bibitem{BibDB:YangKavcic}
S.~Yang and A.~Kav\v{c}i\'{c}, ``Markov sources achieve feedback capacity of
  finite-state machine channels,'' in \emph{Proc. {IEEE} Int. Symp. Inform.
  Theory ({ISIT})}, Lausanne, Switzerland, June 2002, p. 361.

\bibitem{BibDB:ShamaiLaroia}
\mbox{S. Shamai (Shitz)} and R.~Laroia, ``The intersymbol interference channel:
  lower bounds on capacity and precoding loss,'' vol.~42, no.~5, pp.
  1388--1404, Sept. 1998.

\bibitem{BibDB:ArnoldLoeligerIT}
D.~Arnold, H.~A. Loeliger, P.~O. Vontobel, A.~Kav\v{c}i\'{c}, and W.~Zeng,
  ``Simulation-based computation of information rates for channels with
  memory,'' submitted for publication.

\bibitem{BibDB:BCJR}
L.~R. Bahl, J.~Cocke, F.~Jelinek, and J.~Raviv, ``Optimal decoding of linear
  codes for minimizing symbol error rate,'' vol.~20, no.~3, pp. 284--287, Mar.
  1974.

\bibitem{BibDB:ChenSiegel}
J.~Chen and P.~H. Siegel, ``On the symmetric information rate of
  two-dimensional finite state \mbox{ISI} channels,'' in \emph{Proc. {IEEE}
  Information Theory Workshop ({ITW})}, Paris, France, Mar. 2003.

\bibitem{BibDB:Shental}
O.~Shental, N.~Shental, A.~J. Weiss, and Y.~Weiss, ``Generalized belief
  propagation receiver for near-optimal detection of two-dimensional channels
  with memory,'' in \emph{Proc. {IEEE} Information Theory Workshop ({ITW})},
  San Antonio, Texas, USA, Oct. 2004.

\bibitem{BibDB:Tanaka}
T.~Tanaka, ``A statistical-mechanics approach to large-system analysis of cdma
  multiuser detectors,'' vol.~48, pp. 2888--2910, Nov. 2002.

\bibitem{BibDB:BookCover}
T.~M. Cover and J.~A. Thomas, \emph{Elements of Information Theory}.\hskip 1em
  plus 0.5em minus 0.4em\relax John Wiley and Sons, 1991.

\bibitem{BibDB:Leroux}
B.~G. Leroux, ``Maximum-likelihood estimation for hidden markov models,''
  \emph{Stochastic Processes and their Applications}, vol.~40, pp. 127--143,
  1992.

\bibitem{BibDB:BookMezardEtAl}
M.~M\'{e}zard, G.~Parisi, and M.~A. Virasoro, \emph{Spin Glass Theory and
  Beyond}.\hskip 1em plus 0.5em minus 0.4em\relax Singapore: World Scientific
  Lecture Notes in Physics Vol. 9, 1987.

\bibitem{BibDB:ReportYedidiaEtAl}
\BIBentryALTinterwordspacing
J.~S. Yedidia, W.~T. Freeman, and Y.~Weiss, ``Constructing free energy
  approximations and generalized belief propagation algorithm,'' Mitsubishi
  Electric Laboratories, Cambridge, MA, Tech. Rep. TR-2004-40, May 2004.
  [Online]. Available: \url{http://www.merl.com}
\BIBentrySTDinterwordspacing

\end{thebibliography}

\end{document}